\documentclass[11pt,twoside]{article}
\usepackage{./asp2014}
\usepackage{graphicx}
\newcommand*{\vcenteredhbox}[1]{\begingroup
\setbox0=\hbox{#1}\parbox{\wd0}{\box0}\endgroup}
\def\ltsim{\mathrel{\hbox{\rlap{\hbox{\lower4pt\hbox{$\sim$}}}\hbox{$<$}}}}
\newcommand{\degr}{$^{\circ}$}

\aspSuppressVolSlug
\resetcounters

\begin{document}

\title{Review: Magnetic fields of O-type stars}
\author{G.A. Wade$^1$ and the MiMeS Collaboration 
\affil{$^1$Department of Physics, Royal Military College of Canada, Kingston, ON, Canada K7K 7B4 \email{wade-g@rmc.ca}}
}

% This section is for ADS Processing.  There must be one line per author.
\paperauthor{Gregg A. Wade}{wade-g@rmc.ca}{}{Royal Military College of Canada}{Department of Physics}{Kingston}{Ontario}{K7K 7B4}{Canada}

\begin{abstract}
Since 2002, strong, organized magnetic fields have been firmly detected at the surfaces of about 10 Galactic O-type stars. In this paper I will review the characteristics of the inferred fields of individual stars, as well as the overall population. I will discuss the extension of the 'magnetic desert', first inferred among the A-type stars, to O stars up to 60 solar masses. I will discuss the interaction of the winds of the magnetic stars with the fields above their surfaces, generating complex 'dynamical magnetosphere' structures detected in optical and UV lines, and in X-ray lines and continuum. Finally, I will discuss the detection of a small number of variable O stars in the LMC and SMC that exhibit spectral characteristics analogous to the known Galactic magnetic stars, and that almost certainly represent the first known examples of extragalactic magnetic stars.
\end{abstract}

\section{Introduction}
O-type stars are the most massive and luminous stars. Due to their intense UV luminosities, dense and powerful stellar winds, and short lifetimes, they exert a disproportionate impact on the structure, chemical enrichment and evolution of galaxies. 

Through their influence on rotation and mass loss, stellar magnetic fields are able to modify the evolution of O stars in important ways. By coupling to stellar winds, magnetic fields significantly enhance the shedding of rotational angular momentum, leading to rapid spindown and very slow rotation. Organized magnetic fields quench mass loss by trapping wind plasma in regions of closed field lines, producing an accumulation that eventually returns to the stellar surface. Strong magnetic fields may also modify internal differential rotation and circulation currents, changing the transport of both angular momentum and nucleosynthesis products.

O-type stars are progenitors of neutron stars and stellar-mass black holes. The rotation of the cores of red supergiants \citep{2014ApJ...793..123M}, the characteristics of core collapse supernova explosions \citep{2005ApJ...626..350H}, and the relative numbers, rotational properties and magnetic characteristics of neutron stars (and their exotic component of magnetars) may be sensitive to the magnetic properties of their O-type progenitors. The most massive O-type stars have also been associated with the origin of long-soft gamma-ray bursts. 

Considering the importance of O stars as drivers of galactic structure and evolution, and the significance of magnetic fields in determining their evolution, understanding the magnetic characteristics of O stars has been a subject of considerable interest for about a decade.

In this paper we review the historical processes by which magnetic fields were first detected in O-type stars. We summarise the general properties of the currently-known sample of magnetic O stars, emphasizing the importance of the "Of?p" spectral classification. We discuss the surface chemical properties of the magnetic O stars.  To place the magnetic sample in a broader context, we discuss the upper limits for fields in the "non-magnetic" population, and the significance and implications of this apparent dichotomy. We review the detailed characteristics of a few stars of particular importance. We then review recent studies aimed at understanding the physical interaction between the magnetic field and the stellar wind which imprints the optical spectra of magnetic O stars. We discuss the influence of the wind-field interaction on the ultraviolet spectra of O stars, as well as their X-ray emission. Finally, we discuss recently-discovered Of?p stars in the Magellanic Clouds, and evaluate the likelihood that these represent the first positively-identified extra-Galactic magnetic stars.

\begin{figure}
\centering
\vcenteredhbox{\includegraphics[width=2.4in]{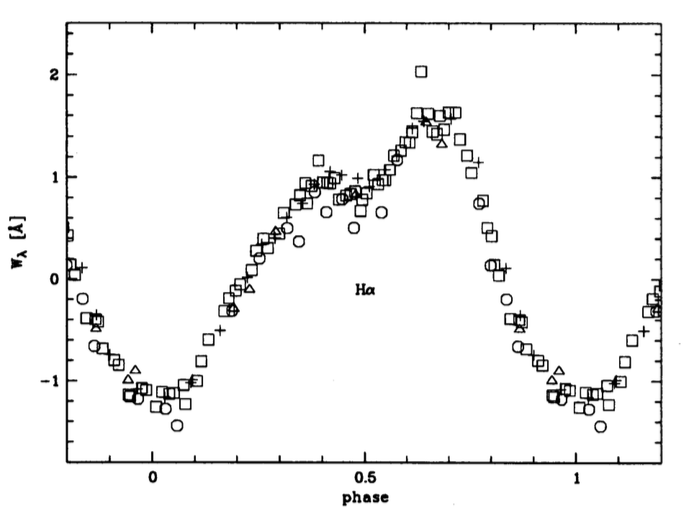}}\hspace{0.5cm}\vcenteredhbox{\includegraphics[width=2.0in]{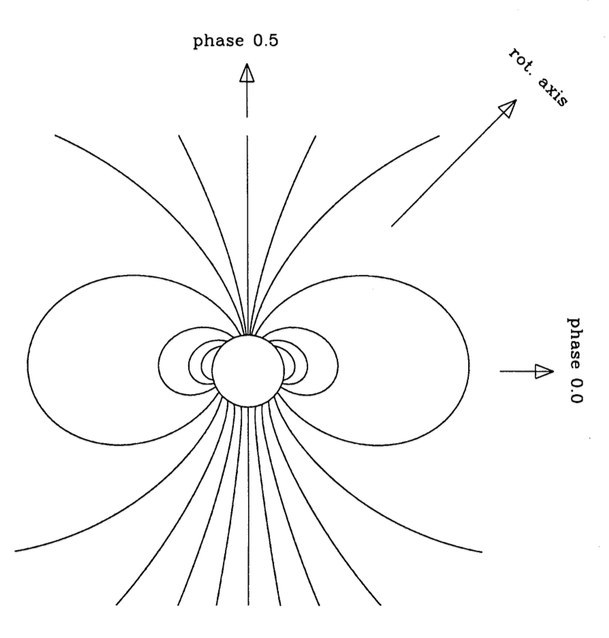}}
\caption{Inference of a magnetic field in the photosphere and wind of the young O-type star $\theta^1$~Ori C by \citet{1996A&A...312..539S}. {\em Left -}\ Equivalent width variation of the H$\alpha$ emission line, phased with the inferred stellar rotational period of 15.4~d. The measurements were acquired during more than 3 years, corresponding to more than 60 cycles. {\em Right -}\ Illustration of the oblique, dipolar magnetic geometry proposed by \citet{1996A&A...312..539S}. In their model, the dipole is inclined to the stellar rotational axis by 45\degr. In their model, the magnetic equator is presented to the observer at phase 0.0 (i.e. at maximum emission), whereas a magnetic pole is visible at phase 0.5 (minimum emission).}
\label{tetori}
\end{figure}

\section{Indirect evidence of magnetism in O-type stars}

\begin{figure}
\centering
\vcenteredhbox{\includegraphics[width=3in]{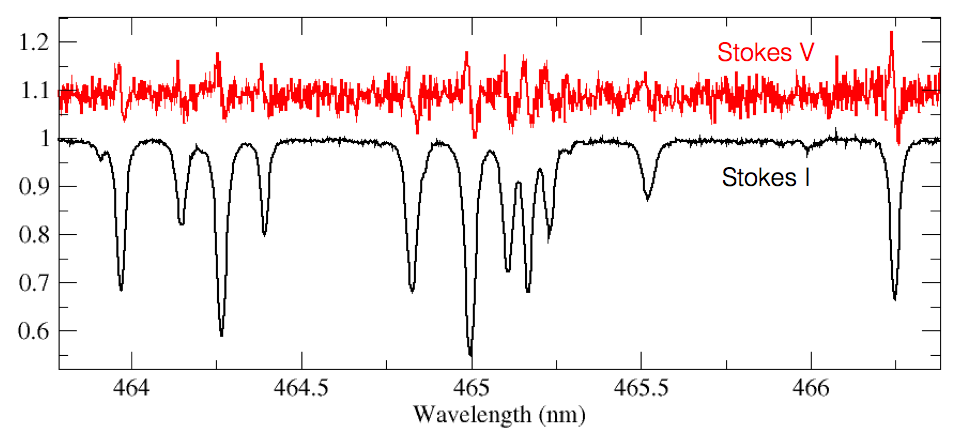}}\vcenteredhbox{\includegraphics[width=2in]{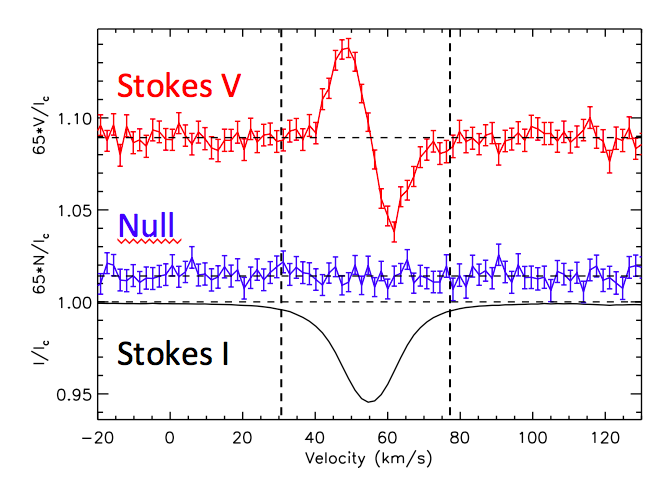}}
\caption{Measurement of stellar magnetic fields using high resolution circular polarization spectroscopy. {\em Left -}\ A small region of a circularly polarized spectrum of the magnetic B0.5V star HD~63425. The lower (black) curve shows the unpolarized (Stokes $I$) spectrum, while the upper (red) curve shows the circular polarization (Stokes $V$) spectrum. Note the weak polarization variation across each spectral line - the signature of the presence of a photospheric magnetic field. {\em Right -}\ The mean Stokes $I$ and $V$ line profiles of HD~63425 computed using the multiline method Least-Squares Deconvolution \citep{1997MNRAS.291..658D,2010A&A...524A...5K}. Note the improvement in the significance of the magnetic detection, which in this case corresponds to a longitudinal magnetic field of $130\pm 10$~G.}
\label{magnetic}
\end{figure}

Ubiquitous structure and variability of the UV wind-sensitive spectral lines of O stars prompted the proposal that magnetic fields, possibly similar to those known to modulate the winds of some B-type stars \citep[e.g.][]{1978ApJ...224L...5L}, were present in the photospheres and winds of those objects \citep[e.g.][]{1994Ap&SS.221..115K}. Although recent observations and modelling have ruled out the general presence of organized magnetic fields capable of producing the observed phenomena \citep{2014MNRAS.444..429D}, it was the particular variability of the outstanding star $\theta^1$~Ori C that led to the first detection of a stellar magnetic field in an O-type star.

$\theta^1$~Ori C is a young O7V star, and the principal ionizing source of the Orion nebula. \citet{1996A&A...312..539S} reported a long time series of optical and UV spectroscopy of this star, all of which demonstrated modulation of wind-sensitive, and possibly photospheric, diagnostic lines with a unique and stable period of 15.4d. As illustrated in Fig.~\ref{tetori}, equivalent width measurements of the H$\alpha$ emission line, obtained during more than 3 years, phase coherently with this period. 

After examining and rejecting a number of hypotheses aimed at explaining the observed periodic variability, \citet{1996A&A...312..539S} proposed that $\theta^1$~Ori C is an oblique magnetic rotator hosting a strong, approximately dipolar magnetic field embedded in the its photosphere. The magnetic field channels the outflowing stellar wind, breaking the spherical symmetry of the stellar mass loss. The axis of the magnetic field is tilted significantly ($\sim 45$\degr) relative to the rotation axis, leading to modulation of wind-sensitive spectral lines as different regions of the magnetically-channelled wind are brought into the view of an observer due to rotation. In this framework, the 15.4d period corresponds to the rotational period of the star. The proposed geometry is illustrated in Fig.~\ref{tetori}.

\begin{table}[!ht]
\smallskip
\begin{center}
{\small
\begin{tabular}{l l l c c llllllll }  % l = left, c = centered
\tableline
%\noalign{\smallskip}\begin{table}{l l l c c D{*}{\,$\pm$\,}{4,4} D{*}{\,$\pm$\,}{4,4} D{*}{\,$\pm$\,}{6,6} c c c c D{*}{\,}{2,4} } 
%\begin{tabular}
%\hline
{ID} &  {Spec. type} &  {$T_{\rm eff}$} & {$\log g$} & {$M$} & {$P_{\rm rot}$} & {$v\sin i$} & {$B_{\rm p}$} \\
 { } & { } & {(kK)} & {cgs}  & {($M_\odot$)} & {(d)} & {(km/s)} & {(kG)} \\
%{(1)} & {(2)} & {(3)} & {(4)} & {(5)} & {(6)} & {(7)} & {(8)} & {(9)} & {(10)} & {(11)} & {(12)} & {(13)} \\
\tableline
 HD\,148937    & O6\,f?p &  41$\pm$2& 4.0$\pm$0.1 &   60 & 7.0323 & $<$~45 & 1.0 \\ 

CPD\,-28\degr\,2561   & O6.5\,f?p &  35$\pm$2& 4.0$\pm$0.2  &  43 & 73.41 &  & >1.7 \\ 

 HD\,37022 &  O7\,Vp &39$\pm$1& 4.1$\pm$0.1 &   45 & 15.424 & 24 & 1.1 \\ 

 HD\,191612 &    O6\,f?p&35$\pm$1& 3.5$\pm$0.1 &  30 & 537.2 & $<$~60 & 2.5 \\ 

 NGC\,1624-2 &    O6.5\,f?cp &  35$\pm$2& 4.0$\pm$0.2 &   34 & 158.0 & $<$~3 & >20 \\ 

 HD\,47129 &  O7.5\,III & 33$\pm$2& 4.1$\pm$0.1  & 56 & 1.21  & 305 & >2.8 \\ 

 HD\,108 &    O8\,f?p &  35$\pm$2& 3.5$\pm$0.2 &   43 & 18000 & $<$~50 & >0.50 \\ 

Tr16-22 &  O8.5\,V &  34$\pm$2 & 4.0$\pm$0.2 &   28 & 54.4  & 25 & >1.5 \\ 

 HD\,57682 &    O9\,V &  34$\pm$1& 4.0$\pm$0.2  &  17 & 63.571 & 15 & 1.7 \\ 

 HD\,37742	 & O9.5\,Ib & 29$\pm$1& 3.2$\pm$0.1  &  40 & 7.0 & 110 & >0.070 \\ 
 
 HD\,54879 & O9.7\,V & & & & & 6-8 & $>$2\\

\noalign{\smallskip}
\tableline\
\end{tabular}
}
\caption{Summary of the properties of known magnetic O-type stars. Adapted from \citet{2013MNRAS.429..398P}. For HD\,54879, no physical parameters have been reported. The $v\sin i$ is adopted from \citet{2014A&A...562A.135S}, and the dipole strength from \citet{2014arXiv1408.2100M}. The period for Tr16-22 is reported by \citet{2014A&A...569A..70N}.}
\label{summary}
\end{center}
\end{table}

Meanwhile, \citet{1997ApJ...478L..87G} detected variable X-rays from $\theta^1$~Ori C that appeared to be modulated with the 15.4d period of Stahl et al. \citet{1997ApJ...485L..29B}, leveraging their Magnetically-Confined Wind Shock (MCWS) model developed to explain X-ray emission and variability from the magnetic Bp star IQ Aur \citep{1997A&A...323..121B}, attempted to understand the X-ray variability in the context of the oblique magnetic rotator. Their model quantified the basic phenomenological picture of Stahl et al., and in particular provided a quantitative theoretical prediction of the presence and intensity of the dipolar surface magnetic field. 

The MCWS model explains the X-ray emission as the consequence of hot plasma, heated by large-scale shocks as wind flows from opposite magnetic hemispheres are driven together at speeds of order the wind terminal velocity ($\sim 1000$~km/s). The resultant equilibrium circumstellar structure consists of a large magnetically-confined volume of shock-heated gas, along with a region of cooler, denser plasma located approximately in the magnetic equatorial plane. Variability of the X-ray flux was explained by periodic occultation of the hot gas by the optically-thick cooling disc. Recent and ongoing studies of $\theta^1$~Ori~C based on modern X-ray data (Cohen et al., in prep.) continue to support the general validity of the MCWS framework.

\section{Detection of magnetic fields in O-type stars and the magnetic field of $\theta^1$~Ori C}

Notwithstanding attempts to directly validate the picture of Stahl et al. and the predictions of Babel \& Montmerle \citep{1999A&A...341..216D}, the magnetic field of $\theta^1$~Ori C remained hypothetical until the detection by \citet{2002MNRAS.333...55D} of Stokes $V$ Zeeman signatures in the averaged optical line profiles of the star. This direct detection of the magnetic field represented the first measurement of a magnetic field in an O-type star.

Stellar magnetic fields are most often detected and characterized using measurements of small polarization variations (particularly circular polarization) across photospheric spectral lines. In the presence of an external magnetic field, atomic electronic transitions become split in a manner that is sensitive to both the field strength and its local orientation. Circular polarization (Stokes $V$), which is produced as a consequence of a line-of-sight, or longitudinal, component of the magnetic field, represents the most sensitive field diagnostic. As illustrated in Fig.~\ref{magnetic} (left panel), the presence of a magnetic field in the stellar photosphere produces characteristic signatures ("Zeeman signatures") with similar shapes, but different amplitudes, in many spectral lines. Modern multi-line methods such as the widely-used Least-Squares Deconvolution \citep[LSD; ][]{1997MNRAS.291..658D,2010A&A...524A...5K} exploit this self-similarity to co-add profiles, improving the signal-to-noise ratio significantly (Fig.~\ref{magnetic}, right panel).

\begin{figure}
\centering
\vcenteredhbox{\includegraphics[width=2.2in]{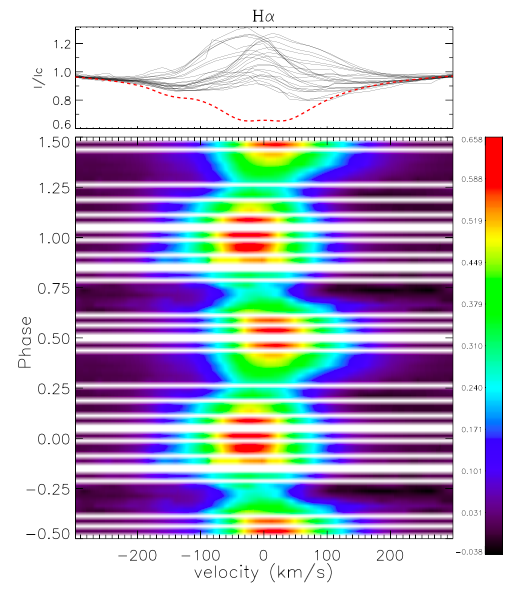}}\vcenteredhbox{\includegraphics[width=2.8in]{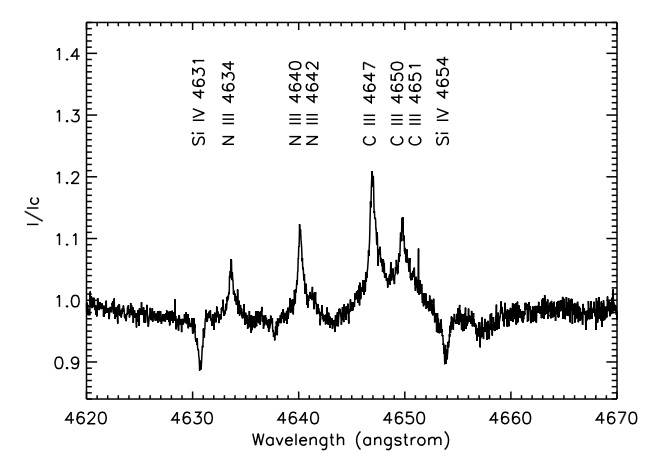}}
\caption{Optical spectral characteristics of magnetic O stars. {\em Left -}\ Dynamic spectrum of the H$\alpha$ profile variability of the magnetic O9IV star HD 57682, with individual profiles overplotted at the top \citep{2012MNRAS.426.2208G}. {\em Right -}\ Of?p diagnostic emission lines near maximum emission in the spectrum of the Of?cp star NGC 1624-2 \citep{2012MNRAS.425.1278W}.}
\label{spectral}
\end{figure}

Based on 5 measurements of the longitudinal magnetic field of $\theta^1$~Ori C roughly sampling the 15.4d period, Donati et al. demonstrated the variability of the longitudinal field component (as expected from a rotating dipole), and inferred that the dipole axis was inclined to the stellar rotation axis by 42\degr. They also inferred that the polar strength of the magnetic field, at the dipole's pole, was 1.1~kG. Notably, the inferred field strength was a factor of $\sim 3$ times larger than that predicted by the MCWS model, and the magnetic variation was out of phase by 180\degr\ relative to the Stahl et al. model. \citet{2006A&A...451..195W} obtained additional circular polarization observations, confirming the field detection and firmly establishing the 15.4d longitudinal magnetic field modulation. More recent observations using more modern instrumentation \citep[e.g.][]{2008MNRAS.387L..23P,2008A&A...490..793H} continue to support the basic characteristics of the magnetic field of $\theta^1$~Ori C as derived by \citet{2002MNRAS.333...55D} and as hypothesized by \citet{1996A&A...312..539S}.

\section{General characteristics of magnetic O-type stars}
\label{characteristics}

As summarized in Table~\ref{summary}, there are currently 11 established magnetic O-type stars. Most of these stars exhibit physical, magnetic and rotational properties analogous to those of $\theta^1$~Ori C: they have masses ranging from $\sim 15-50~M_\odot$, their inferred magnetic field strengths are typically of order 1-3~kG, and they rotate slowly as compared to non-magnetic O stars ($P_{\rm rot}$ typically in the range from 1 week to several months, but extending to years/decades). Ten of the magnetic O stars listed in Table~\ref{summary} were discussed by \citet{2013MNRAS.429..398P}. An additional object, the late-type star HD 54879, was recently added by the BoB collaboration \citep{2014arXiv1408.2100M}.

\subsection{Optical spectral characteristics}
\label{sectsummary}

All known magnetic O stars show complex optical spectra, typically containing numerous emission lines. Most lines are variable, and wind-sensitive lines are usually strongly variable. When sufficient observations exist, it is demonstrated that variability occurs according to a unique, well-defined and stable period that is identified as the stellar rotational period (as in the case of $\theta^1$~Ori C). The H$\alpha$ line, as well as the He~{\sc i}~$\lambda 5876$ and He~{\sc ii}~$\lambda 4686$ lines, are the principal features used to diagnose and interpret the variability. Variability of the H$\alpha$ line profile of a late-type magnetic O star, HD 57682, is illustrated in Fig.~\ref{spectral} (left panel).

Of the known magnetic O stars, 5 are associated with the peculiar spectral classification "Of?p". This classification was first introduced by \citet{1972AJ.....77..312W} according to the presence of C~{\sc iii} $\lambda 4650$ emission with a strength comparable to the neighbouring N~{\sc iii} lines (see Fig.~\ref{spectral}, right panel). Well-studied Of?p stars are now known to exhibit recurrent, and apparently periodic, spectral variations (in Balmer, He~{\sc i}, C~{\sc iii} and Si~{\sc iii} lines), narrow P Cygni or emission components in the Balmer lines and He~{\sc i} lines, and UV wind lines weaker than those of typical Of supergiants \citep[see][and references therein]{2010A&A...520A..59N}.

Only 5 Galactic Of?p stars are known \citep{Walbetal10a}: HD 108, HD 148937, HD 191612, NGC 1624-2 and CPD\, -28\degr 2561. All of these have been carefully examined for the presence of magnetic fields \citep[][Wade et al., submitted]{2006MNRAS.365L...6D,2010MNRAS.407.1423M,2011MNRAS.416.3160W,2012MNRAS.419.2459W,2012MNRAS.425.1278W,2011A&A...528A.151H} and all have been clearly detected. It therefore appears that the particular spectral peculiarities that define the Of?p classification are a consequence of their magnetism.

\begin{figure}
\centering
\vcenteredhbox{\includegraphics[width=2.52in]{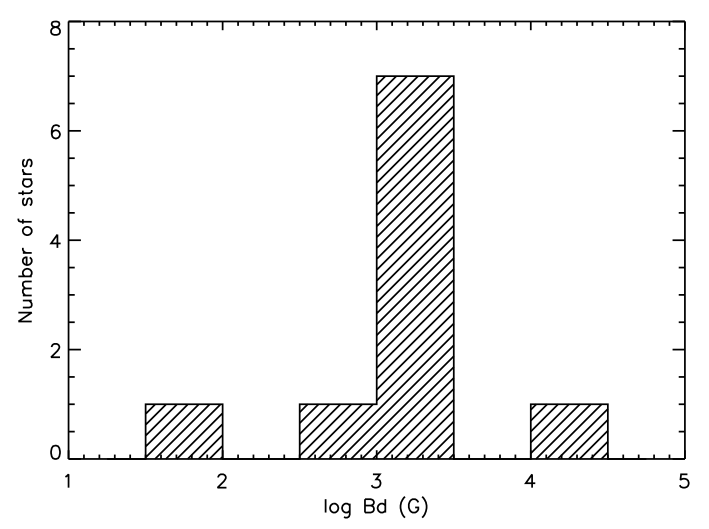}}\vcenteredhbox{\includegraphics[width=2.5in]{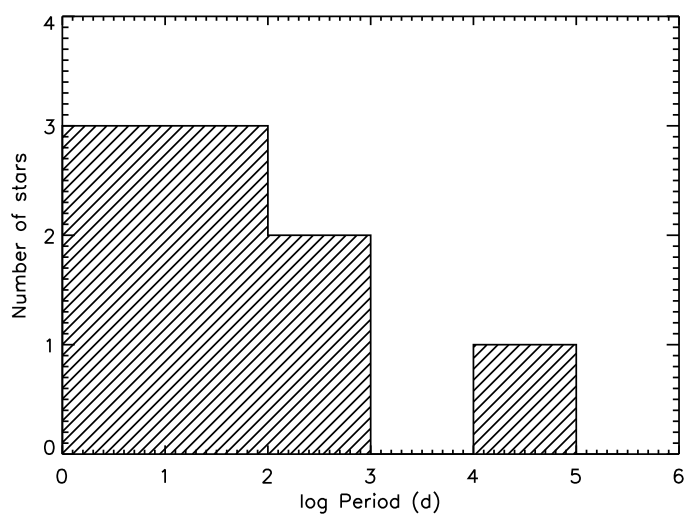}}
\caption{Distributions of dipole strengths and rotational periods of magnetic O stars. {\em Left -}\ Dipole strengths. {\rm Right -}\ Rotational periods.}
\label{histograms}
\end{figure}

\subsection{Magnetic fields and rotation}

As illustrated in Fig.~\ref{histograms} (left panel), the inferred surface dipole strengths of magnetic O stars are typically between 1-3 kG. However, field strengths exhibit a significant range, from a few hundred gauss (an upper limit in the case of HD 37742=$\zeta$~Ori Aa; Blaz\`ere et al., in preparation) to about 20~kG \citep[NGC 1624-2][]{2012MNRAS.425.1278W}. Due to the complexity of their spectra, the precision and accuracy of models of the surface magnetic fields of O stars are generally poorer than those obtained for cooler (i.e. B and A-type) stars. For example, as illustrated by the analyses of \citet{2012MNRAS.419.2459W} (HD 148937) and Wade et al. (submitted) (CPD\, -28\degr 2561), the magnetic field strength is often inferred from a longitudinal field variation measured with a sigificance of only a few $\sigma$. On the other hand, additional constraint on the magnetic geometry is often available from interpretation of emission line variations \citep[e.g.][Wade et al., submitted]{2012MNRAS.426.2208G}.

The magnetic fields of O stars appear to be stable on timescales of many rotations, i.e. at least months-years. In the case of $\theta^1$~Ori C, magnetic measurements obtained during well over a decade (i.e. over hundreds of rotations) phase together acceptably. These field characteristics are understood in the physical context of the fossil field paradigm, in which the magnetic fields are believed to be the slowly-decaying remnants of field that was accumulated or produced early in the formative history of the stars. Recent theoretical efforts \citep[e.g.][]{2004Natur.431..819B,2010A&A...517A..58D,2010ApJ...724L..34D,2013MNRAS.428.2789B} have provided fundamental support that has served to underpin this framework.

As discussed above, the rotational periods of magnetic O stars are inferred from periodic modulation of emission and absorption line profiles, the longitudinal magnetic field, and sometimes from photometric variability. Periods for 7 of the 11 known magnetic O stars are well established (from the published literature for HD 148937, HD 37022, HD 191612, NGC\, 1624-2, and HD 57682, by Wade et al. (submitted) for CPD\, -28\degr 2561, and by Grunhut et al. (in prep) for HD 47129). For HD 108, the period ($\sim 55$y) is estimated based on the historical record of photometric and spectroscopic variability \citep[e.g.][]{2001A&A...372..195N}. For Tr16-22, a period of $\sim$54d has been proposed from X-ray data alone, but needs to be confirmed in using optical diagnostics \citet{2014A&A...569A..70N}. For HD 37742, the period is estimated qualitatively based on spectroscopic variability. For  HD\,54879, no period has yet been published. As illustrated in Fig.~\ref{histograms} (right panel), the rotational periods of all but one star are at least one week. In comparison with the period of $\sim 4$d for a typical non-magnetic O star with $v\sin i$ of 100~km/s and $R=10~R_\odot$ and $i=60$\degr, rotation of magnetic O stars is systematically slow. In some cases rotational periods are dramatically longer: months (in the cases of HD 57682, NGC 1624-2, Tr16-22 and CPD\, -28\degr 2561), years (in the case of HD 191612), and likely decades (in the case of HD 108). The slow rotation is typically attributed to loss of rotational angular momentum through magnetic coupling to the wind \citep[magnetic braking;][]{2009MNRAS.392.1022U}. It is interesting to note that while rotational periods of magnetic O stars are often remarkably long, in a number of cases periods are only somewhat longer than would be expected for a typical non-magnetic O star (i.e. 7 days  in the case of HD 148937).

A significant outlier in terms of rotational periods is HD 47129 (Plaskett's star). In this SB2 system (discussed in further detail below), it is the rapidly-rotating secondary star in which the magnetic field is detected. Based on both the measured $v\sin i$ \citep[$\sim 300$~km/s;][]{2013MNRAS.428.1686G} and the rotational period inferred through modulation of spectral and magnetic measurements (1.21d; Grunhut et al., in prep.), this star is a rapid rotator. Although \citet{2013MNRAS.428.1686G} explained this characteristic in the context of spin-up due to post-Roche lobe overflow of the primary star, recent observations (Grunhut et al, in prep., and described below) question this interpretation.

 \begin{figure}
\centering
\vcenteredhbox{\includegraphics[width=2.5in]{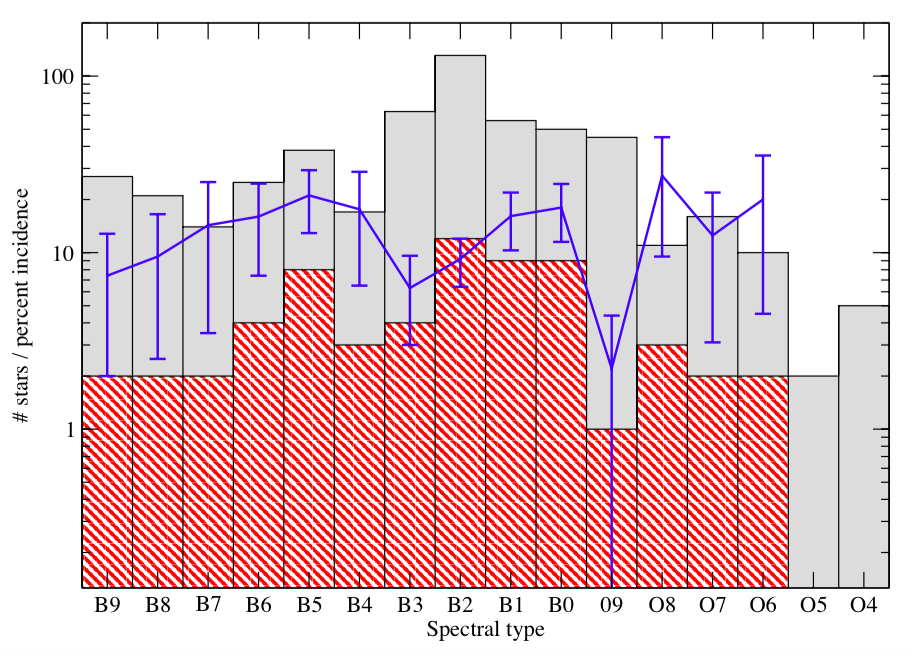}}\vcenteredhbox{\includegraphics[width=2.25in]{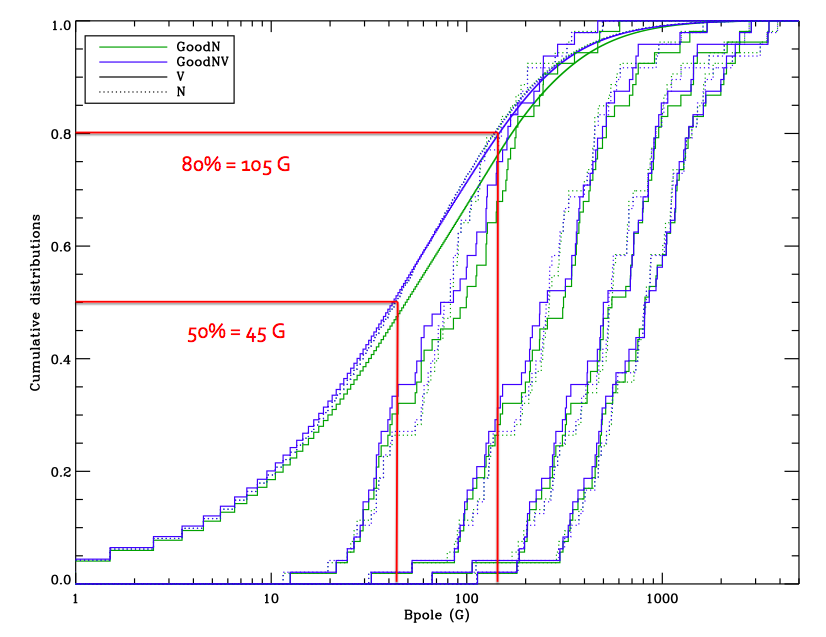}}
\caption{{\em Left -}\ The incidence of magnetic fields in B and O-type stars as inferred from the MiMeS survey. The grey histogram illustrates the number of stars per spectral type bin observed in the survey. The red histogram illustrates the number of stars per spectral type bin detected to be magnetic in the survey. The blue curve (with error) bars illustrates the computed incidence of magnetic stars as a function of spectral type. {\em Right -}\ The leftmost curve shows the field strength distribution built directly from the probability density functions of individual undetected O stars using the approach of \citet{2012MNRAS.420..773P}. The derived median and 80th percentile inferred field strengths are 45~G and 105~G, respectively. The 4 curves on the right side shows the cumulative distributions built from the upper limits of the 68.3, 95.4, 99.0, and 99.7 percent credible regions, for both Stokes $V$ and the diagnostic null. From Petit et al. (in prep.).}
\label{incidence}
\end{figure}

\subsection{Incidence of magnetic fields in O-type stars}

The MiMeS survey of magnetism in massive stars \citep{2014IAUS..302..265W} included Stokes $V$ observations of more than 100 O-type stars. The bulk incidence (i.e. the total number of previously-unknown magnetic stars in the sample relative to the total sample) is $7\pm3$\% (where the uncertainties are computed from counting statistics). The general magnetic characteristics of the B-type stars and O-type stars detected in the survey are very similar.

Analysis of the undetected magnetic O stars by Grunhut et al. (in prep.) yields a median longitudinal field error bar of 44~G, about $2\times$ smaller than the best error bars obtained for the magnetic O star HD 191612. Petit et al. (in prep.) have applied the modelling approach of \citet{2012MNRAS.420..773P} to infer upper limits on the dipole magnetic fields of the undetected sample, taking into account uncertainties on the magnetic geometry and phase in a Bayesian statistical framework. These results imply upper limits on surface dipoles of the sample of 40~G at 50\% confidence, and 130~G at 80\% confidence. This strongly suggests a bimodal, rather than continuous, distribution of magnetic properties of O stars: a small population with strong ($\gtrsim 1$~kG) magnetic fields, and a much larger population with magnetic fields that are at present undetected, and generally smaller than $\sim 100$~G. This is reminiscent of the "magnetic dichotomy" established for A-type stars by \citet{2007A&A...475.1053A}.

We conclude that the incidences and characteristics of large-scale magnetic fields in B and O type stars are indistinguishable based on their observational properties, and qualitatively identical to those of intermediate mass stars with spectral types $\sim$F0 to A0 on the main sequence. The MiMeS survey therefore establishes that the basic physical characteristics of magnetism in stellar radiative zones remains unchanged across more than 1.5 decades of stellar mass, from spectral types F0 ($\sim 1.5~M_\odot$) to O4 ($\sim 50~M_\odot$). 

These results are described in more detail by \citet{2014IAUS..302..265W} and by Grunhut et al. and Petit et al. (in prep). The magnetic incidence as a function of spectral type, and the cumulative distributions related to modelling of dipole fields of the undetected O stars of the MiMeS sample, are illustrated in Fig.~\ref{incidence}.

\begin{figure}
\centering
\vcenteredhbox{\includegraphics[width=4.5in]{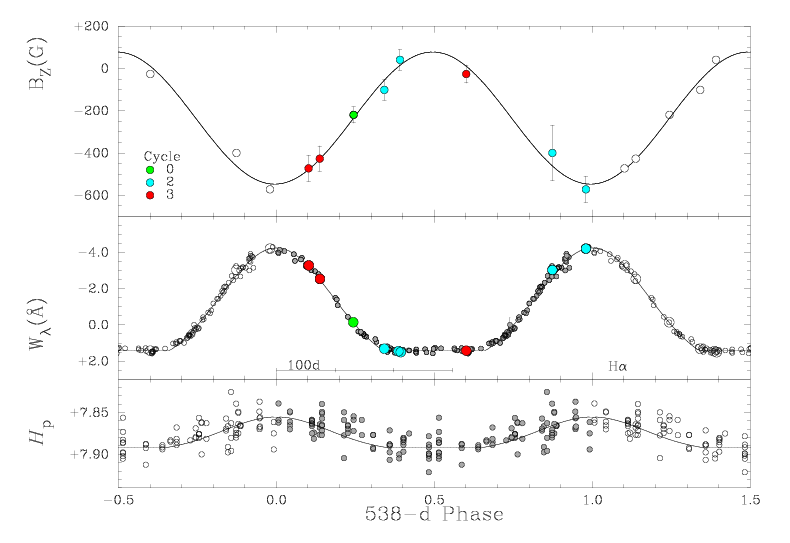}}
\caption{Magnetic, spectral and photometric variations of the magnetic O star HD 191612, phased according to the star's 538d rotational period. {\em Top -}\ Longitudinal magnetic field. {\em Middle -}\ H$\alpha$ equivalent width. {\em Bottom -}\ Hipparcos $H_{\rm p}$ magnitude. From \citet{2011MNRAS.416.3160W}. }
\label{hd191612}
\end{figure}

\subsection{Surface chemistry}

Significant recent discussion and research \citep[e.g.][]{2011A&A...530A.116B,2014ApJ...781...88A} has been focused on understanding the origin of peculiar surface abundances of light elements in B-type stars, and their relationship to magnetic fields.

Investigations of the surface chemical abundances of magnetic O-type stars have been conducted by \citet{2012A&A...538A..29M}, and revisited and placed within the context of a larger sample of non-magnetic O stars by Martins et al. (submitted). These studies show that surface abundances of single O stars exhibit trends with age that are reasonably well explained by current stellar evolution models. The magnetic O stars have surface abundances that do not depart from the main trends of other O-type stars. 

\subsection{Unconfirmed claims}

A number of O-type stars that have been claimed to be magnetic in the literature have been found to be non-magnetic on further examination. Most of these claims were based on low-significance detections obtained with the FORS1 and FORS2 instruments at VLT. The majority are discussed by \citet{2012A&A...538A.129B}, who attribute these false detections to important contributions of noise from non-photon sources such as small instrument flexures. Magnetic detections that have been claimed for O-type stars that do not appear in Table~\ref{summary} should be interpreted with care.

 \section{Individual magnetic O-type stars of note}
 \label{individual}
 
 \subsection{HD~191612: a first test of the oblique rotator model for magnetic O stars}
 
 HD 191612 is an Of?p star and the second O-type star in which a magnetic field was detected \citep{2006MNRAS.365L...6D}. Long-term investigations of its variability from the optical to X-rays by \citet{2004ApJ...617L..61W}, \citet{2007MNRAS.375..145N} and \citet{2007MNRAS.381..433H} established that all observational indicators varied with a single, well-defined period of 538d. In particular, they demonstrated that this variability was distinct from the binary radial velocity variation with a period of 1548d.
 
\citet{2011MNRAS.416.3160W} obtained magnetic field observations spanning over 4 years, and demonstrated that the longitudinal magnetic field also varied according to the 538d period. As illustrated in Fig.~\ref{hd191612}, the magnetic field, H$\alpha$ equivalent width and Hipparcos magnitude phase coherently according to the spectral period, and exhibit a clear phase relationship in which the extrema of all three quantities occur at the same phases.
 
 The H$\alpha$ equivalent width measurements were obtained over a timespan of nearly 30 years, i.e. $\sim20$ cycles. While the large-scale variation of the profile is stable and periodic, the H$\alpha$ profiles are not strictly repeatable: the dispersion of equivalent width measurements about the mean variation is somewhat larger than the observational uncertainties \citep{2007MNRAS.381..433H}. This "jitter", which appears to occur on timescales longer than a few days, is observed more clearly in the case of the Of?p star HD 148937 \citep{2012MNRAS.419.2459W}.
 
 \citet{2011MNRAS.416.3160W} interpreted the combined variations of HD 191612 in the context of the magnetic oblique rotator model, as proposed by \citet{1996A&A...312..539S} for $\theta^1$~Ori C. The fundamental principle of this model is that the variability is driven by or associated with the tilted magnetic field, and that the variability period is the stellar rotational period. Their reasoning relied essentially on the demonstration that the magnetic field variation occurs with the same period as the spectral and photometric variations, with a clear phase relationship between these quantities. 
 
 Studies of the relationship between projected rotation velocity and variability period of magnetic A-type stars have demonstrated the validity of this model in that context beyond all reasonable doubt. However, the important macroturbulent broadening of the line profiles of many O-type stars precludes such a test of the model. For example, in the case of HD 191612, the upper limit on $v\sin i$ obtained from modelling line profiles is 60~km/s, whereas the predicted rotational speed according to the oblique rotator model is $<1.5$~km/s. Some cooler magnetic O stars \citep[e.g.][]{2012MNRAS.426.2208G} provide a limited capability to test the model in this way. A further check could be performed using observations of linear polarization due to scattering from the magnetically-confined wind \citep[][]{2012MNRAS.419.2459W, 2013ApJ...766L...9C}.
 
 \subsection{NGC 1624-2: a magnetic O star with an extraordinarily strong field}
 
\begin{figure}
\centering
\vcenteredhbox{\includegraphics[width=5in]{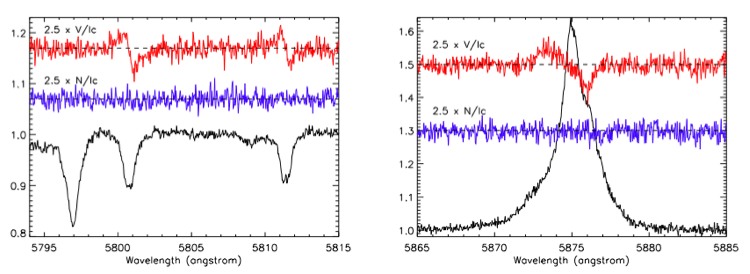}}
\caption{Stokes $I$ and $V$ line profiles of NGC 1624-2, illustrating the detection of strong Zeeman signatures. From \citet{2012MNRAS.425.1278W}. }
\label{dash2-1}
\end{figure}

NGC 1624-2 is the faintest ($V=11.8$) of the known Galactic Of?p stars. \citet{2012MNRAS.425.1278W} demonstrated that the star varies with a well-defined period of 158d, which they interpreted as the stellar rotational period. They reported the detection of a very strong magnetic field in this star - sufficiently strong to produce Zeeman signatures (Fig.~\ref{dash2-1}) in individual spectral lines that are detectable in spectra of modest quality ($<100$ SNR). Those authors also reported the detection of probable resolved Zeeman splitting in the unpolarized profiles of a few spectral lines. The measured longitudinal magnetic field of $>5$~kG, along with the line splitting, provide direct evidence of a surface magnetic field with a dipole component of at least 20 kG. Such a field is nearly one order of magnitude stronger than those detected in other O stars.

\begin{figure}
\centering
\vcenteredhbox{\includegraphics[width=2.3in]{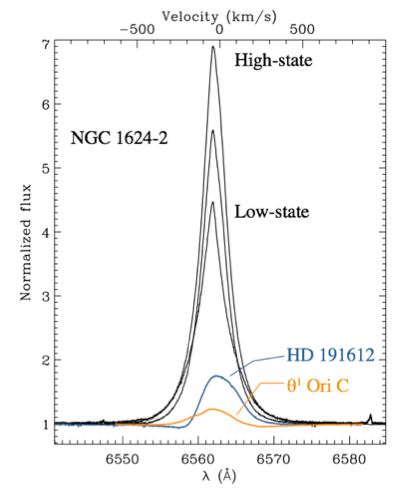}}\hspace{0.5cm}\vcenteredhbox{\includegraphics[width=2.5in]{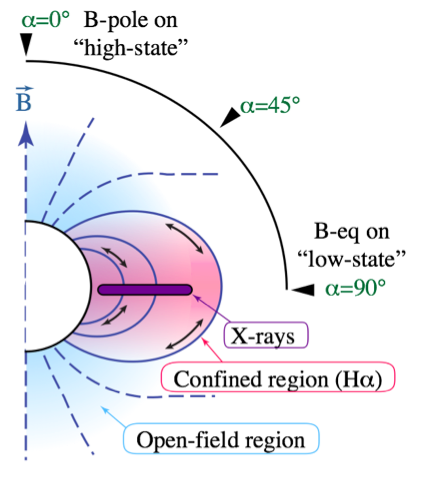}}
\caption{{\em Left -}\ H$\alpha$ profiles of NGC 1624-2 at maximum, intermediate and low emission, compared to maximum emission profiles of HD 191612 and $\theta^1$~Ori C. Adapted from \citet{2012MNRAS.425.1278W}. {\em Right -}\ Illustration of the magnetospheric structure and emission regions of NGC 1624-2 and other magnetic O stars. Arrows at top right indicate the viewing directions of observation at phases 0.0 ($\alpha=0$\degr), 0.25 ($\alpha=45$\degr) and 0.5 ($\alpha=90$\degr). These figures are provided courtesy of V. Petit.}
\label{dash2-2}
\end{figure}

There are several consequences of such a strong field. First, the field appears capable of stabilizing the stellar atmosphere down to great depths \citep{2013MNRAS.433.2497S}, significantly modifying the dynamical behaviour of the gas. This is discussed further below. Secondly, the intense magnetic field is capable of controlling the flow of the wind to very large distances from the star (see Sect.~\ref{interaction}), generating an immense stellar magnetosphere. Evidence for huge quantities of cool, confined plasma is provided by the strong H$\alpha$ emission of NGC 1624-2, illustrated in Fig.~\ref{dash2-2} (left panel). \citet{2012MNRAS.425.1278W} calculate that the closed magnetosphere of NGC 1624-2 (Fig.~\ref{dash2-2}, right panel) extends to a distance of $11\pm 4~R_*$, nearly 3 times larger than any other magnetic O star, and implying a volume of confined plasma that is more than an order of magnitude greater. 

NGC 1624-2 is also an intense X-ray source. As illustrated in Fig.~\ref{dash2-2} (right panel), in the context of Babel \& Montmerle's MCWS model, the large-scale wind collision occurring near the magnetic equatorial plane produces shock heating of the gas, yielding X-ray emission that may be modulated by rotation. As reported by \citet{2012MNRAS.425.1278W,2014arXiv1409.1690N}, while the X-ray efficiency is rather typical of magnetic O-type stars (about $4\times$ greater than for non-magnetic O stars), X-rays appear to be harder than for most other magnetic O stars. New Chandra observations have revealed a strong, phase-dependent attenuation of the X-ray emission which suggest a much larger magnetospheric absorption than in any other magnetic O-type star (Petit et al. in prep).

\citet{2013MNRAS.433.2497S} examined the optical absorption lines of NGC 1624-2 and compared them to other magnetic O-type stars and normal O-stars. They noticed that the spectral lines of NGC 1624-2 were very narrow - in fact, that they could be reproduced by considering only the broadening due to the star's strong magnetic field. In contrast, the spectral lines of other magnetic O stars required an important macroturbulent broadening to explain their large line widths under the constraint of their slow rotation. In other words, the line profiles of most magnetic O stars were similar to those of non-magnetic O stars, whereas the lines of NGC 1624-2 were much narrower. \citet{2013MNRAS.433.2497S} demonstrated that the much stronger field of NGC 1624-2 was capable of freezing motions in the photosphere to very deep layers - in fact, layers corresponding to the Fe convection zone recently discussed in the literature as the possible origin of photospheric macroturbulence. They suggested that the narrow lines supported the proposal that macrotubulent broadening of O star line profiles has its origin in their deep subphotospheric layers.

 \subsection{HD 47129: a rapidly-rotating magnetic O star in a short-period binary?}
 
 \begin{figure}
\centering
\vcenteredhbox{\includegraphics[width=4.5in]{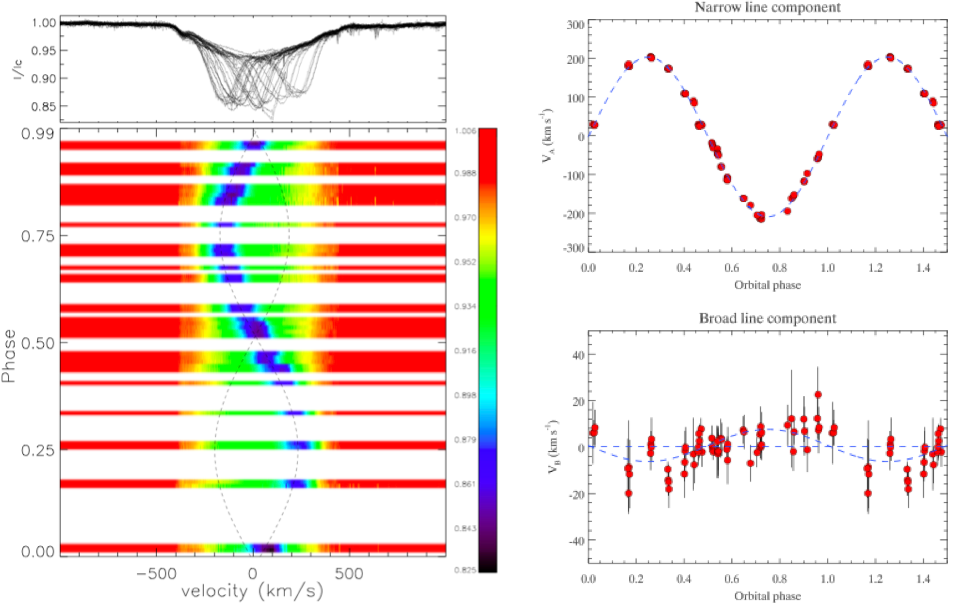}}
\caption{Variations of the He~{\sc Ii} $\lambda 5411$ line in the spectrum of HD 47129. {\em Left -}\ Dynamic spectrum of the combined lines of the primary and secondary star, with spectra phased according to the orbital period. Whereas the primary's narrow line is observed to vary sinusoidally in radial velocity with a full amplitude of $\sim 400$~km/s, the secondary's line is motionless. The expected RV motion of the two stars is illustrated by the dashed curves. An overplot of the line profiles is shown at the top. {\rm Right -}\ Measured radial velocities of the primary and secondary star from spectral disentangling. In both cases the best-fit sinusoidal variation is shown. Note the difference in vertical axis scales. From Grunhut et al. (in prep.).}
\label{plaskett}
\end{figure}
 
 HD 47129 (Plaskett's Star) has historically been understood to be a massive binary system comprised of two $\sim 50~M_\odot$ O-type components, orbiting a common centre-of-mass with a period of about 14.5d \citep[e.g.][]{2008A&A...489..713L}. The system is known to be a hard, luminous and variable X-ray emitter \citep[e.g.][]{2006MNRAS.370.1623L}, and has been interpreted as a colliding-wind binary (CWB) \citep{1992ApJ...396..238W}. The components of HD 47129 are also reported to exhibit peculiar surface chemistry, very different projected rotation velocities (70~km/s for the primary, 305~km/s for the secondary), and a mass-luminosity mismatch (in the sense that the fainter secondary star is inferred to have a mass superior to the primary).
 
 \citet{2013MNRAS.428.1686G} reported the detection of a magnetic field in the broad-lined secondary component of the system. The inferred magnetic field strength and topology ($\sim 2.5$~kG, dipolar) are similar to those of other well-studied magnetic O stars. However, the very high projected rotation velocity - 305~km/s - is at odds with the typically slow rotation of such stars. In fact,  \citet{2013MNRAS.428.1686G} derived a rotational period of the magnetic star of $1.7\pm 0.5$d, and tentatively associated the photometric period of 1.22d detected in CoRoT photometry by \citet{2011A&A...525A.101M} with the rotation of the secondary.  Such rapid rotating stands in stark contrast with the typical rotation periods ($\sim$months) of other magnetic O stars.
 
It has been proposed that the peculiar physical, rotational and chemical characteristics of HD 47129 can be explained if is a post mass transfer (i.e. post RLOF) system \citep{2008A&A...489..713L}. Such a scenario also naturally attributes the rapid rotation of the secondary to spin-up due to mass transfer, and neatly explains the absence of slow rotation. Nevertheless, an additional problem has come to light based on continued analysis of this system by Grunhut et al. (in prep.) in the context of the BinaMIcS project. 

As illustrated in Fig.~\ref{plaskett}, recent measurements of the radial velocities of the two components of HD 47129 from high signal-to-noise ratio, high resolution spectra demonstrate that while the radial velocity of the primary star varies sinusoidally with a full amplitude of $\sim 400$~km/s, the velocity of the secondary is stable. This stands in stark contrast to previously-reported variations of the secondary's radial velocity of hundreds of km/s. As explained by Grunhut et al. (in prep.), this is likely explained by sensitivity of the employed disentangling methods to the initial assumed orbital solution.

These results have important implications for the interpretation of the HD 47129 system, since it is not immediately clear how to reconcile the behaviour of the line profiles with a dynamically bound system.

 \section{The wind-magnetic field interaction}
 \label{interaction}

 Systematic MHD studies of the interaction of outflowing radiatively-driven stellar winds with dipolar magnetic fields have been carried out during the past decade \citep[e.g.][]{2002ApJ...576..413U,2008MNRAS.385...97U,2009MNRAS.392.1022U,2013MNRAS.428.2723U}. A basic conclusion of these investigations is that two physical parameters are capable of describing the general behaviour of the wind of a hot star under the influence of a magnetic field and stellar rotation \citep{2013MNRAS.429..398P}: the {\em wind magnetic confinement parameter} \citep{2002ApJ...576..413U} $\eta_*=B_{\rm eq}^2R^2/{\dot M} v_\infty$ (which measures the energy density in the magnetic field relative to the wind kinetic energy density) and the {\em rotation parameter} \citep{2008MNRAS.385...97U} $W=V_{\rm rot}/V_{\rm orb}$ (which measures the equatorial rotational velocity of the star relative to its breakup speed). These parameters serve to respectively define the Alfv\'en radius $R_{\rm A}$ (the radius corresponding to the approximate extent of the last closed magnetic loops in the magnetic equatorial plane) and the Kepler co-rotation radius $R_{\rm K}$ (at which the Keplerian orbital period is equal to the stellar rotational period). 
 
 In the case of a rapidly-rotating star, the Kepler radius is located relatively close to the stellar surface, and for sufficiently strong magnetic fields is located inside the Alfv\'en radius. In this scenario, plasma in the region between $R_{\rm K}$ and $R_{\rm A}$ is forced (by the magnetic field) to orbit at greater than the local Keplerian speed, and hence experiences an unbalanced (outward) net  force. In such a "centrifugal magnetosphere", wind plasma is trapped in this region by the combined effects of magnetic field and rotation. In the case of a slowly-rotating star, the Kepler radius is located far from the stellar surface. The net gravitational + centrifugal force is always directed toward the star. In such a "dynamical magnetosphere" scenario, plasma driven up the field lines ultimately cools and falls back to the stellar surface.
 
 As described in Sect.~\ref{characteristics}, most magnetic O stars are slow rotators; hence their magnetized winds are "dynamical magnetospheres". 
 
\begin{figure}
\centering
\vcenteredhbox{\includegraphics[width=4in]{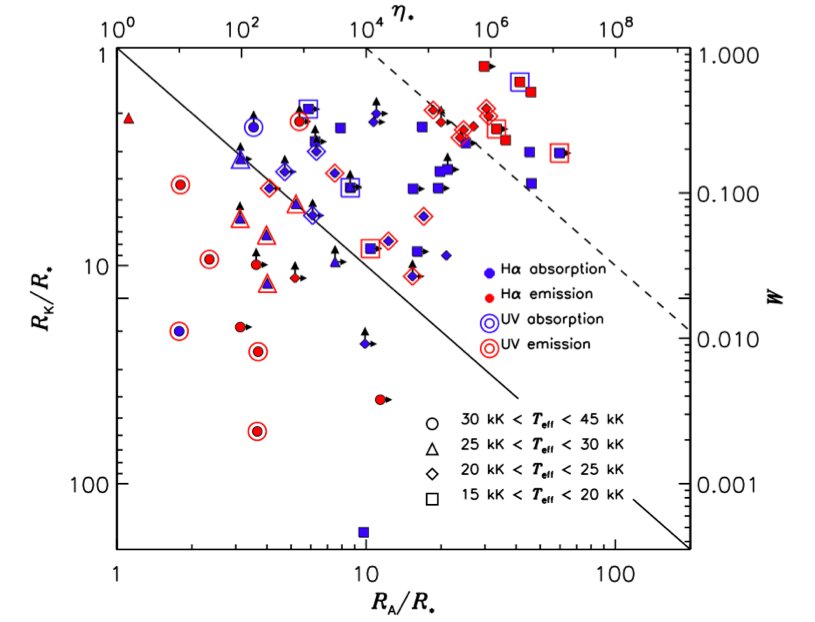}}
\caption{The rotation-confinement diagram summarizing the characteristics of magnetized winds of OB stars \citep[][Shultz et al. 2014]{2013MNRAS.429..398P}. Note that most magnetic O stars (indicated by circles) are located at the left of the diagram, reflecting their slow rotation and their classification as "dynamical magnetospheres".}
\label{rotcon}
\end{figure}

 These principles were summarized by \citet{2013MNRAS.429..398P} and developed into a two-parameter "rotation-confinement" diagram. As illustrated in Fig.~\ref{rotcon} (from Shultz et al., these proceedings) the O type stars are all located at the left of the diagram, reflecting their slow rotation and moderate wind confinement.
 
 Current 2D and 3D MHD models can effectively compute the evolution of the wind under such conditions, and have provided a sound theoretical basis for understanding the general observational behaviour (e.g. as described in Sect.~\ref{individual}) of magnetic O stars.

\subsection{Modeling the H$\alpha$ line}
\label{secthalpha}
 
\begin{figure}
\centering
\vcenteredhbox{\includegraphics[width=2.25in]{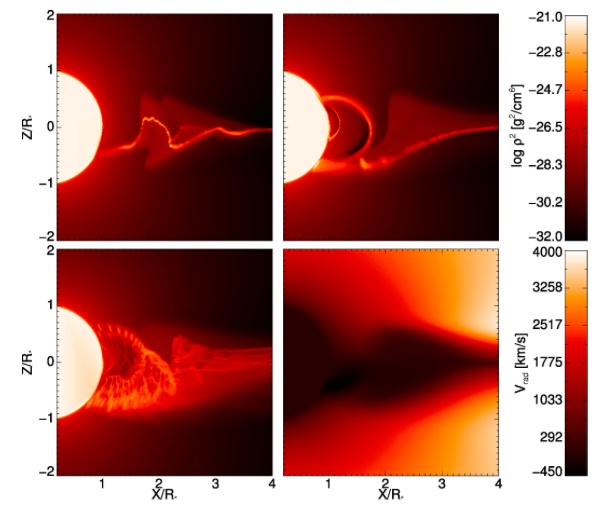}}\vcenteredhbox{\includegraphics[width=2.55in]{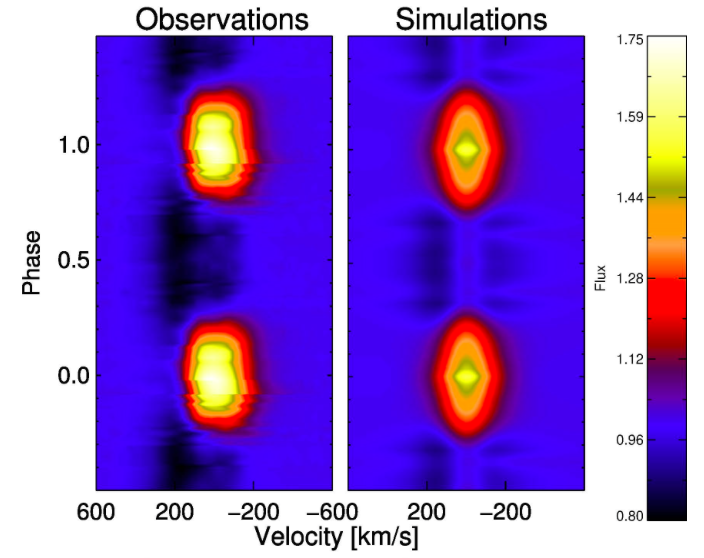}}
\caption{{\em Left -}\ Contours of the squared density for two different snapshots of the 2D MHD wind simulation employed for the synthesis of the H$\alpha$ profile of HD 191612 and its variability. (upper panels). Time-averaged density-squared (lower left) and radial velocity (lower right). {\em Right -}\ Observed and synthetic H$\alpha$ dynamic spectra of HD 191612, as a function of rotation phase. The model spectra have been convolved with an isotropic macroturbulence of 100 km/s. From \citet{2012MNRAS.423L..21S}.}
\label{halpha}
\end{figure}

\citet{2012MNRAS.423L..21S} presented a modelling approach to understand the H$\alpha$ line variations of magnetic O-type stars. They employed a time average of over 100 snapshots of the density, temperature and velocity fields computed in 2D MHD simulations of the evolving magnetically-confined wind of HD 191612 to construct a pseudo-3D wind model using an "orange-slice" approach (Fig.~\ref{halpha}, left panel). Adopting a stellar and magnetic geometry consistent with those determined observationally, they computed the predicted H$\alpha$ profile variation versus rotational phase by solving the formal integral of radiative transfer in a 3-D cylindrical coordinate system. They were able to reproduce reasonably well the phase variation of the H$\alpha$ line (both its equivalent width and dynamic spectrum: Fig.~\ref{halpha}, right panel), although the computed profile was significantly narrower than the observations. This was resolved by convolving the computed profiles with a Gaussian corresponding to a macroturbulent velocity of 100~km/s. \citet{2012MNRAS.423L..21S} suggested that this required macroturbulence is an artefact of missing wind dynamics in the 2D simulations, and would likely we resolved by full MHD simulations in 3D. 

The 2D MHD simulations employed by \citet{2012MNRAS.423L..21S} \citep[and by][and Wade et al., (submitted) in subsequent applications of this approach]{2012MNRAS.426.2208G} assumed an initial dipolar magnetic field with its axis of symmetry aligned with the stellar rotation axis. Because of the slow rotation of the stars considered, it was argued that the impact of rotation on the dynamics of the wind could be neglected, and so that such models were considered to be fully appropriate.

Recently, \citet{2013MNRAS.428.2723U} have succeeded in computing 3D MHD simulations of the radiatively-driven wind of a magnetic O star ($\theta^1$~Ori C). These new simulations, still computed only for rotation aligned with the magnetic field axis, accurately reproduce the H$\alpha$ equivalent width variation and effective validate the results from 2D simulations of \citet{2012MNRAS.423L..21S}. However, the discrepancy between the widths of the observed and computed profiles remains. Simulations including the effects of an inclined magnetic field may serve to resolve them. This would indicate that aligned-field models are missing physics that is important in reproducing the complete range of characteristics of the observed line profiles.

 \subsection{UV spectroscopy}
 
Detailed HST Space Telescope Imaging Spectrograph (STIS) UV spectra of two magnetic O-type stars (HD 191612 and HD 108) have been acquired and analysed, and are discussed in the literature \citep{2012MNRAS.422.2314M,2013MNRAS.431.2253M}. In addition, a new STIS dataset of the Of?p star CPD\,-28\degr 2561 has recently been acquired by Naz\'e et al., and observations of NGC 1624-2 are currently scheduled.

 \begin{figure}
\centering
\vcenteredhbox{\includegraphics[width=3.5in]{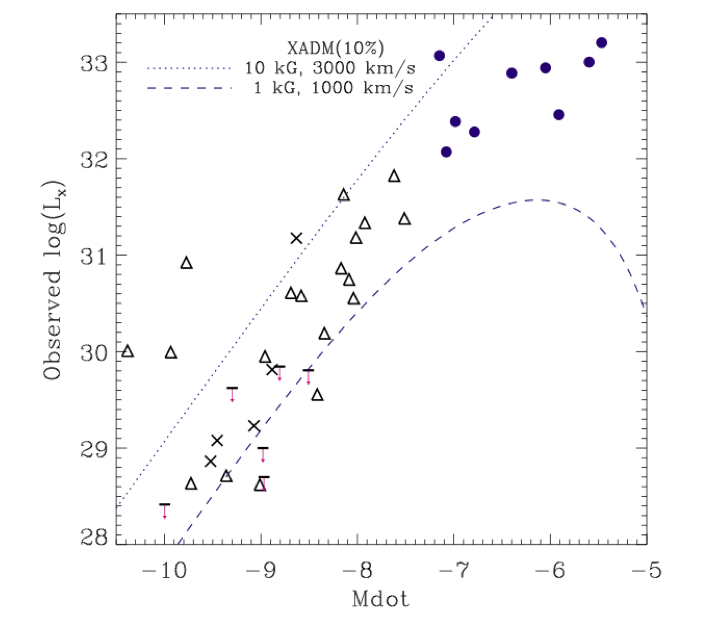}}
\caption{Scaling of X-ray luminosity with mass-loss rate \citep{2014arXiv1409.1690N}. Magnetic O-type stars are indicated by filled circles. The dotted and dashed curves represent the predictions of the XADM model for two different magnetic field/wind parameter combinations. Note that the model successfully predicts the lower slope of the $L_{\rm X}-\dot M$ power-law relation at high mass-loss rate.}
\label{xrays}
\end{figure}

\citet{2013MNRAS.431.2253M} observed that UV spectra of HD 191612 observed at different rotational phases were broadly similar, with the same set of wind and photospheric lines consistently present. However, the line profile variations were very significant compared to those of normal O stars. Nevertheless, the variations are more subtle than those observed in optical lines such as H$\alpha$, likely due to the greater distance from the star at which the UV lines are formed. They demonstrated that the equivalent widths of UV spectral lines of HD 191612 varied coherently when phased with the 538d period derived from optical spectroscopy, in a manner consistent with the oblique rotator model. Comparison of the data with the predictions of spectral synthesis assuming a 1D spherically symmetric wind yielded acceptable overall agreement; however, details of the line profiles, as well as the phase variability, were not reproduced. A remarkable discovery was the anti-correlated behaviour of the (strong) C~{\sc iv} lines and the (weak) Si~{\sc iv} resonance lines. They used quasi-3D radiation MHD wind simulations, similar to those employed by \citet{2012MNRAS.423L..21S} in Sect.~\ref{secthalpha}, to synthesise the variable profiles of UV scattering lines using the '3D SEI' method. These models reproduce the anti-correlated behaviour of the C and Si lines, qualitatively explaining this difference as due to different sampling of the magnetically-confined plasma and the approximately unperturbed wind by saturated and unsaturated lines. 

Results for HD 108 by \citet{2012MNRAS.422.2314M} are broadly compatible with the conclusions of \citet{2013MNRAS.431.2253M}. In analogy with HD 191612, the UV spectroscopic results for HD 108 (which leveraged a previously IUE spectrum obtained at the star's 'high state' in addition to the STIS data obtained at the star's 'low state') support the interpretation of \citet{2010MNRAS.407.1423M} that HD 108 is a magnetic rotator with a very long rotational period \citep[$\sim 55$y; ][]{2010A&A...520A..59N}.
 
\subsection{X-rays}
 
Following sporadic X-ray studies of individual magnetic O stars \citep[e.g.][]{2005ApJ...628..986G,2012ApJ...746..142N}, a systematic investigation of the X-ray properties of magnetic OB stars has recently been carried out by \citet{2014arXiv1409.1690N}. Those authors collected all available Chandra and XMM-Newton exposures of known massive magnetic stars, corresponding to over 100 observations including $\sim 60$\% of stars compiled in the catalogue of \citet{2013MNRAS.429..398P}. Overall, they found that the X-ray luminosity of magnetic massive stars is strongly correlated with the stellar wind mass-loss rate, with a power-law form that is slightly steeper than linear for the majority of B stars, and flattens for the O stars. No significant dependence of any other X-ray characteristic (e.g. hardness, inferred local absorption) on other parameters (e.g. magnetic field strength) was observed. They concluded that the observed X-ray luminosities, and their trends with mass-loss rates, could not be reproduced by the 'embedded wind shocks' paradigm that explains the X-ray emission from single, non-magnetic O stars. They were also able to successfully reproduce the main trend of X-ray luminosity with mass-loss rate using a scaled version of the semi-analytic X-ray Analytic Dynamical Magnetosphere (XADM) model \citep{2014MNRAS.441.3600U} for X-ray emission from closed magnetic loops.

The scaling of X-ray luminosity with mass-loss rate for the sample of magnetic B+O stars investigated by \citet{2014arXiv1409.1690N} is shown in Fig.~\ref{xrays}.

\section{Of?p stars in the SMC and LMC}
 
\begin{figure}
\centering
\vcenteredhbox{\includegraphics[width=5in]{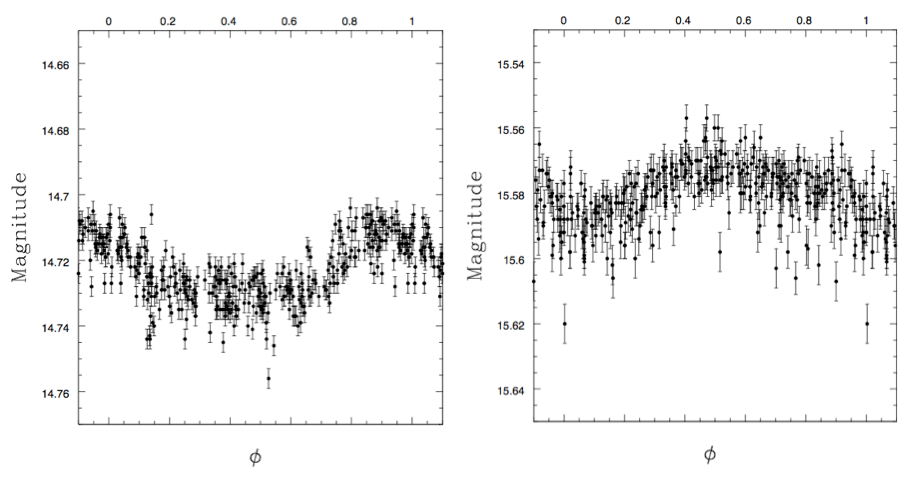}}
\caption{Phased photometry of two LMC/SMC Of?p stars (Naz\'e et al., in prep.). On the left, data are phased with a period of $\sim 8$d, while on the right, they are phased with a period of $\sim 15$d.}
\label{yael}
\end{figure}

Recent large-scale surveys of the O star content of the Magellanic clouds have identified a small number of stars in the LMC and SMC exhibiting the peculiar C~{\sc iii} $\lambda 4650$ emission and strong He~{\sc ii}~$\lambda 4686$ emission characteristic of Of?p stars \citep[][Howarth et al., in prep.]{2000PASP..112.1243W,2001ApJ...550..713M,2014ApJ...788...83M}. Considering the demonstration in Sect.~\ref{sectsummary} that all Galactic Of?p stars host strong, organized magnetic fields, it is likely that these stars are also magnetic oblique rotators, and represent the first known extra-Galactic magnetic stars. Spectroscopic monitoring (Morrell et al., priv. comm.) reveals that at least two of these objects are strong spectrum variables. Naz\'e (in prep.) have recently examined photometry of two of these stars, and have demonstrated that the photometry can be phased according to periods of $\sim 8$d and $15$d, consistent with rotational periods of the known magnetic O stars. These results are illustrated in Fig.~\ref{yael}.
 
 \section{Summary and conclusions}
 
This paper has summarized the current state of understanding of magnetic fields of O-type stars:

\begin{itemize}
\item As illustrated by the first-discovered magnetic O star, $\theta^1$~Ori C, magnetic O stars can generally be identified by strong periodic variability of optical spectral lines resulting from the influence of the magnetic field on their radiatively-driven winds. In fact, many magnetic O stars exhibit sufficiently distinctive spectral peculiarities that their spectral classification is affected (the 'Of?p' stars).
\item Currently, there exist 11 confirmed magnetic O stars, corresponding to a small fraction (about 7\%) of Galactic O stars observed in the MiMeS survey. Their spectral types range from O9.7 to O6, and their masses from 17-60~$M_\odot$. All but one of these stars are main sequence objects.
\item Their magnetic fields are inferred to be strong ($\sim$kG), organized (dipolar), and inclined relative to the stellar rotation axis. Their variability is interpreted in terms of rotational modulation within the framework of the oblique rotator model.
\item The inferred rotational periods of magnetic O stars are typically very long, from 1 week to decades. The typical periods - of a duration of several months - are significantly longer than those expected for 'normal' non-magnetic O stars. This slow rotation is understood as a consequence of rotational angular momentum loss through the wind resulting from magnetic braking.
\item The abundances of light elements in the photospheres of magnetic O stars are compatible with those of non-magnetic O stars.
\item The spectral peculiarities and variability of magnetic O stars is understood principally as the result of magnetic confinement of wind plasma by the magnetic field. MHD simulations (chiefly in two dimensions) have inspired a basic theoretical framework that allows the classification of the magnetically-confined winds of massive stars into two general categories: "centrifugal magnetospheres" and "dynamical magnetospheres". On account of their slow rotation, the magnetized winds of O stars typically fall in the latter category.
\item Radiation MHD models currently allow the detailed synthesis of a limited number of spectral diagnostics of the wind-magnetic field interaction.
\item UV and X-ray studies of magnetic O stars yield important new constraints on the lower density and higher temperature components of their magnetized winds.
\item Of?p stars discovered in the LMC and SMC likely represent the first known extra-Galactic magnetic stars. However, their magnetic fields have yet to be measured directly.
\end{itemize}
 
The small number of currently-known magnetic O stars makes drawing conclusions regarding their general properties a challenge. The application of more sensitive polarimetric instrumentation - on larger, more powerful optical telescopes, as well as extending into the infrared - is likely to yield a significant expansion of the population in the near future. This is important, because magnetic fields of O stars likely contain important clues informing us about the formation of massive stars, and may help to understand the magnetic characteristics of their descendants, the neutron stars and magnetars.

\section{Acknowledgments}

I would like to explicitly acknowledge to contributions of the following collaborators to work described in this review: R. Barb\'a, D. Cohen, A. David-Uraz, A. Fullerton, J.H. Grunhut, H.F. Henrichs, I.D. Howarth, J. Ma\'iz Apellan\'iz, W. Marcolino, F. Martins, Y. Naz\'e, S. Owocki, V. Petit, M. Shultz, J. Sundqvist, A. ud-Doula, N.R. Walborn.

\bibliography{wade}  % For BibTex

\end{document}